\documentstyle[preprint,aps,epsf,floats]{revtex}
\tighten
\input {amssym.def}
\input {amssym.tex}
\everymath{\rm}
\everydisplay{\rm}

\begin{document}
\draft

\title{Detection and Implications of a Time-reversal Breaking State in
Underdoped Cuprates}

\author{ M.E. Simon and C.M. Varma}
\address{Bell Laboratories, Lucent Technologies \\
Murray Hill, NJ 07974}

\maketitle

\begin{abstract}
We present general symmetry considerations on how a Time-reversal breaking state may
be detected by angle-resolved photoemission using circularly polarized photons 
as has been proposed earlier. Results of recent experiments utilizing the proposal 
in underdoped cuprates are analysed and found to be consistent in their symmetry
and magnitude with a theory of the Copper-Oxides.
 These togather with evidence for a quantum critical 
point and marginal Fermi-liquid properties near optimum doping suggest 
that the essentials of a valid
 microscopic theory of the phenomena in the cuprates may have 
 been found.
\end{abstract}

A major problem in condensed matter physics in the past decade and a 
half has been the search
for a microscopic theory of high temperature superconductivity and associated normal state
anomalies \cite{general}. The normal state properties which presage superconductivity imply the
 inapplicability of the quasiparticle concept and are well-described by the marginal
Fermi-liquid phenomenology \cite{cmv1}. This prescribes scale-invariant fluctuations governed by a
quantum-critical point (QCP). A change of symmetry in the normal state with 
doping is then expected. The crucial question is whether a state with broken symmetry 
indeed exists 
and what is its nature? A microscopic theory based on a general model of the 
Cuprate compounds predicts an
elusive phase which breaks time-reversal symmetry, without changing the translational symmetry 
of the lattice \cite{cmv2}. It is characterised by an ordered pattern 
of currents spontaneously circulating in prescribed patterns in each unit cell in the 
underdoped cuprates. Angle-resolved photoemission experiments using polarised light 
were suggested to detect such a phase \cite{cmv3}. Here the general 
symmetry considerations necessary for the
experiment and its analysis are derived. These results have been used in recent 
experimental work \cite{kaminski} to detect a time-reversal breaking phase in underdoped
cuprates. The experiments are 
analysed here to show that the  symmetry of the effect is characterestic 
of the class of the predicted phases and to rule out some other possibilities.

The current patterns in the ground state predicted by the microscopic theory \cite{cmv2}
are illustrated in Fig. (1).
Both arise from the same microscopic Hamiltonian; one or the other has lower
energy depending on the detailed parameters of the model.
It was suggested that the so called "pseudogap phase" observed in these
 compounds is such a T-breaking phase.
 
The lack of a new translational symmetry in this
 phase makes it very hard to detect. Such a phase can be detected by 
 measuring  the difference in intensity of angle-resolved
  photoemission spectra (ARPES) for right and left circular polarized
  photons \cite{cmv3} in a mono-domain sample and analyzing the symmetry of
the difference. The difference should set in below the characterestic
  pseudogap temperature.\begin{figure}[t] 
\begin{center} 
\leavevmode 
\epsfxsize = 12. truecm 
\epsfbox{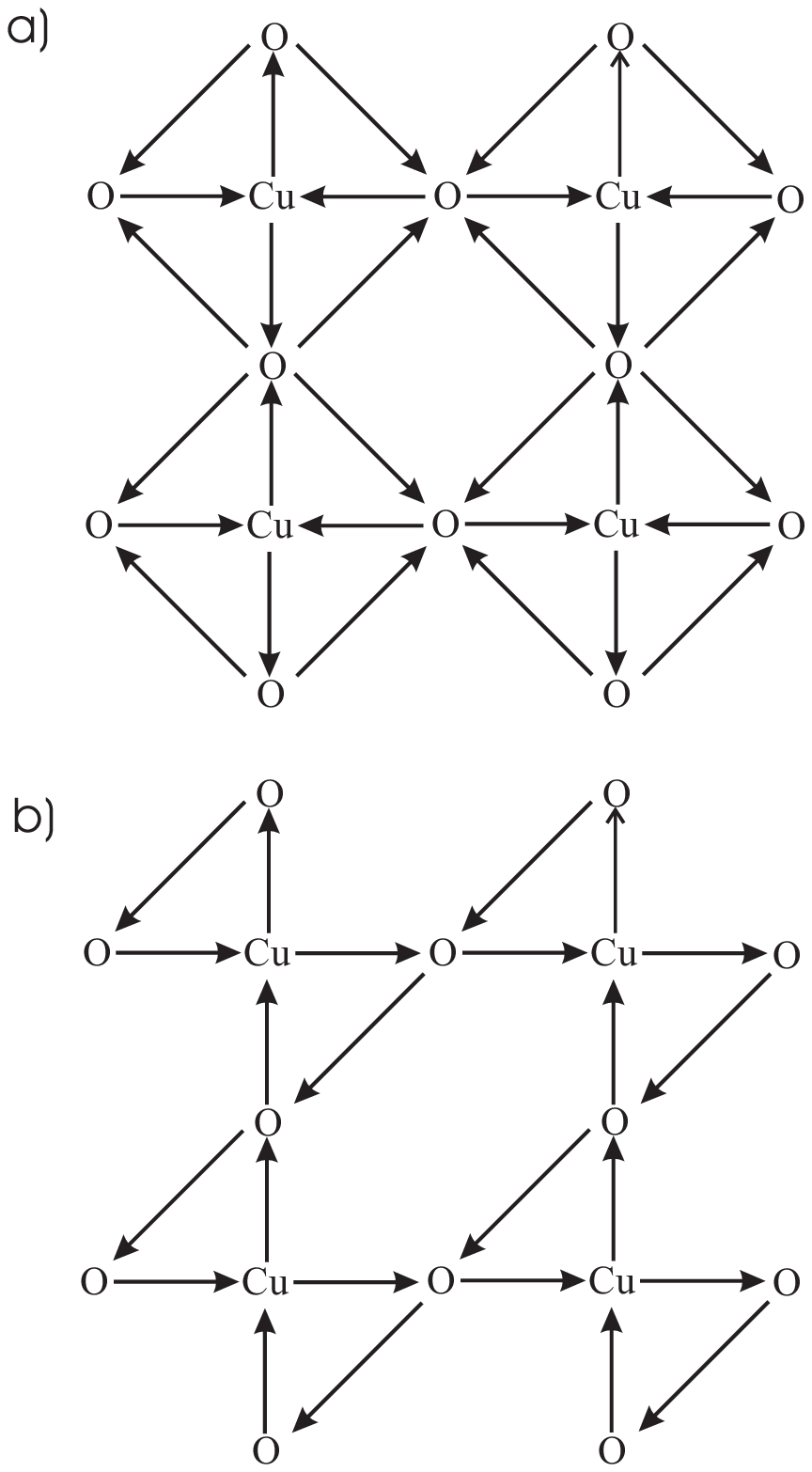}
\end{center}
\caption{ 
Current patterns for the predicted T-breaking states that preserves translational 
invariance} 
\end{figure}

For molecules absorbed on surfaces, a geometric effect has been derived,
   \cite{dubbs} which
   even without T-breaking  yields
  intensity which depends on the circular polarization in ARPES experiments. 
Here we first 
  generalise the geometric effect for the symmetries of
  a crystal and then derive the conditions necessary to distinguish the geometric effect
  from the effect due to T-breaking. Moreover, the symmetry of the experimental results
as the direction of momentum of the outgoing electrons varies is shown to distinguish
between different possible 
T-breaking phases.

{\it General Results}:
  Suppose a beam of photons of energy $\omega$ shone on a crystal in the
  direction $\hat{n}$ produces  free-electons with momentum $\bf {p}$
  and energy $E_ {\bf {p}}$ at the detector. Let $|\bf{k}\rangle$  denote the
  states of the crystal. Here $\bf{k}$ is the wave-vector in the first
  Brillouin zone. Assuming the momentum of the photons is
  small compared to $\bf{k,p}$, the current $J_{\bf{p}}$ is given by
  \begin{equation}
  J_{\bf{p}} = 2\pi e\sum_{\bf{k}}' f(\epsilon_{\bf{k}})
    |\langle {\bf p |M| k}\rangle|^{2} \delta( E_{\bf{p}}-\epsilon_{\bf{k}}+\omega).
    \label{currentequation}
    \end{equation}
where the matrix element  is given by
    \begin{equation}
     \langle {\bf p |M| k}\rangle = \frac{-ie}{2mc}\int d\bf{r} \Phi_{\bf{p}}(\bf {r})
     {\bf{A.\nabla}} \psi_{\bf{k}}(\bf{r}).
     \end{equation}
and the summation is restricted by momentum conservation between $\bf{p,k}$
modulo the reciprocal vectors.
     Also $\bf{A}$ is the vector potential and $\Phi_{\bf{p}} (\bf{r})$ is the
    wave-function of the outgoinh photoelectron
 of momentum $\bf{p}$. We may distinguish the two circular
     polarizations by $\bf{A}_{\ell,r}$
     \begin{equation}
     {\bf {A}}_{\ell,r} =A_0(-\hat{x'}\pm i\hat{y'}),
     \end{equation}
where $\hat{x'}$ and $\hat{y'}$ are perpendicular to $\hat{n}$, and the two matrix elements by ${\bf M}_{\ell,r}(\bf{k,p})$. We will assume that the crystal being studied
is two-dimensional so that $\bf{k}$ refers to the momentum in the x-y plane.  
Note that, since the momentum of the photon is assumed negligible,
$\bf{p} = (\bf{k} + \bf{G})$. So, when $\bf{p}$ is in the mirror plane of the crystal,
so is $\bf{k}$. (The converse is not true.)
 Let $\hat{m}$ be the set of mirror planes
of the crystal normal to the surface of the crystal. For reasons that will be clear shortly,
we will consider only the situation in which $\hat{n}$ lies in one of the $\hat{m}$-planes.
     
      We may write in general that
      \begin{equation}
      |\bf{p}\rangle = \alpha_m |{\bf{p}},e\rangle + \beta_m|{\bf{p}},o\rangle
      \end{equation}
      where under reflection $\Re$ about a given m-plane, 
 \begin{equation}
      \Re_m |{\bf{p}}\rangle = \alpha_m |{\bf{p}},e\rangle - \beta_m |{\bf{p}},o\rangle.
\end{equation}
 In Eq. (4), the eigenstates are divided into two parts,
one of which has a real space representation even in reflection and the other has a real space representation odd in reflection about the given 
$\hat{m}$-plane. 
       Also because $\hat{n}$ is contained in the mirror plane $\hat{m}$ 
 \begin{equation}
      \Re^{-1}_m(\bf{A_{\ell}.\nabla})\Re_m = (\bf{A_r.\nabla})
 \end{equation}
 
     Consider the group of the crystal wavefuctions $|\bf{k} \rangle$.
     In general, we may write,
     \begin{equation}
      |{\bf{k}} \rangle = \mu_m |{\bf{k}},e \rangle + \nu_m |{\bf{k}},o \rangle
      \end{equation}
      so that in reflection about the $\hat{m}$-plane,
      \begin{equation}
       \Re_m|{\bf{k}} \rangle = \mu_m |{\bf{k}},e \rangle -  \nu_m|{\bf{k}},o \rangle.
      \end{equation}
In Eq. (7), the division of the wavefunctions follows the same convention 
as in Eq. (4). When T-symmetry is preserved,
 $\nu_m=0$ if ${\bf{k}}$ lies in the mirror plane

        We are now finally ready to relate the matrix element
         for left circular polarisation (lcp)
        with that for right circular polarization (rcp). 
        Using Eq. (6), we can write,
        \begin{equation}
      {\bf M}_{\ell} =  \langle {\bf{p}}|\Re^{-1} ({\bf{A}_r.\nabla})\Re |{\bf{k}}\rangle.
      \end{equation}
Then using Eqs. (5) and (8),
      \begin{eqnarray}
      {\cal{D}}_m \equiv |{\bf M}_{\ell}|^{2}- |{\bf M}_r|^{2} & = &
        4 {\cal{R}} \left( \alpha_m^*\beta_m |\mu_m|^2  \langle {\bf{p}},e|{\bf M}_r^*|{\bf{k}},e \rangle
\langle {\bf{k}},e|{\bf M}_r |{\bf{p}},o \rangle  \right. \\ \nonumber & + &
      \beta_m^*\alpha_m|\nu_m|^2\langle {\bf{p}},o|{\bf M}_r^*|{\bf{k}},o\rangle \langle {\bf{k}},o|{\bf M}_r|{\bf{p}},e\rangle  \\ \nonumber & + &
\mu_m\nu_m^*|\alpha_m^2\langle {\bf{p}},e|{\bf M}_r^*|{\bf{k}},e\rangle 
          \langle {\bf{k}},o|{\bf M}_r|{\bf{p}},e\rangle \\ \nonumber & + &
     \left. \nu_m\mu_m^*\|\beta_m|^2\langle {\bf{p}},o|{\bf M}_r^*|{\bf{k}},o\rangle
 \langle {\bf{k}},e|{\bf M}_r|{\bf{p}},o\rangle\right).
    \end{eqnarray}
The difference in $\bf{J_p}$ due to rcp and lcp follows through Eq. (1).
In Eq. (10), ${\cal{R}}$ picks up only the real part of its argument. 
We now separately consider the cases, T-symmetry preserved and T-symmetry broken.

{\it T-Symmetry preserved}
As mentioned above a finite $\cal{D}$ exists even in this case due to the geometry of the experiment. 
For a T-preserving hamiltonian
in a crystal with center of inversion $\alpha,\beta,\mu,\nu$ may
be taken real.  
For the geometric effect to be non-zero, it is necessary that either  the
state $|\bf{p}\rangle$
      or the state $|\bf{k}\rangle$,
      does not have definite parity under the indicated reflection; this requires that  
three of the four quantities 
      $\alpha,\beta,\mu,\nu$ are non-zero.
      Since if $\bf{p}$ lies in a mirror-plane, so does $\bf{k}$, the former
      ensures the latter. Thus the geometric effect is zero if $\bf{p}$
lies in the plane $\hat{m}$. Note that we assumed above that 
      $\hat{n}$ lies in the mirror plane $\hat{m}$. If the experimental geometry is such that
$\hat{n}$ does not lie in the plane $\hat{m}$, it is possible to show that the geometric effect is 
present in general even if  $\bf{p}$ lies in the plane $\hat{m}$. (The geometric effect
is also zero, even if the above condition is satisfied, 
if the matrix elements in the product (10) are zero due to
some other symmetries.)

       The induced geometric effect
      must be distinguished in experiments from the proposed effect due to
      T-breaking.  Towards this end, an important result following
      from Eq.(10)
      is that the geometric effect is odd with respect to reflection of $\bf{p}$
      about the m-plane. Thus if the outgoing plane wave with momentum 
$\bf{p}$ has a component
      $\delta \bf{p}_{perp}$ normal to this plane, (i.e. when $\beta \ne 0$,
 the difference of the intensity
      for (rcp)-ARPES and (lcp)-ARPES changes sign for
      $\delta \bf{p}_{perp}~ \rightarrow ~-\delta \bf{p}_{perp}$.]

  {\it T-symmetry Broken}:  A complex hamiltonian breaks T-symmetry if no unitary 
transformation can convert it to a real form. Correspondingly, its eigenstates 
cannot be transformed to a real form by any gauge transformation.
 In translationally invariant media, broken T-symmetry 
implies broken inversion symmetry  (provided
Charge conjugation symmetry exists). In crystalline solids 
this is not necessary. A broken
T-symmetry may or may not imply a broken reflection symmetry about some
crystalline mirror plane \cite{footnote}. Since inversion is a product 
of reflections about three mutually orthogonal mirror
planes, inversion symmetry may be preserved while T-symmetry is broken.
The specific proposals for T-symmetry breaking in copper-oxide metals 
that have been considered
all lead to a broken reflection symmetry about one or more of the 
{\it crystalline} mirror planes

The broken reflection symmetry about a given mirror plane $\bar{m}$ 
attending a broken T-symmetry must be distinguished from that due to
 a structural or electronic distortion.
In the latter cases, diffraction experiments, sensitive to density variations, 
detect the effect.
For broken T-symmetry alone, the charge density retains the reflection symmetry
 about $\bar{m}$ while the
wavefunctions may not. (Specific examples of this will be given below)
In that case we will continue to call $\bar{m}$ a mirror plane. For example
 in the copper-oxide lattice the $x=0,y=0$ and $x=\pm y$ will continue to
 be called mirror planes even though due to broken T-symmetry, the wavefunctions
may not be eigenstates of $\Re$ about one or more of these planes.

This has the following consequence in Eq. (10). Consider $\bf{p}$ in a mirror plane
 $\bar{m}$,
so that $|\bf{p}\rangle=|\bf{p}_e\rangle$. Then although ${\bf k}={\bf p}$ lies in the 
plane $\bar{m}$
the wave-function $|{\bf{k}}\rangle$ has besides the usual component $|\bf{k},e\rangle$,
 a component
$\theta |{\bf k},o\rangle$. It then follows that the third term in Eq. (10) is not zero 
for ${\bf p},{\bf k}$ in the plane $\bar{m}$. This is true only for the 
mirror planes $\bar{m}$ about which reflection symmetry is broken due to T-breaking.
It then also follows that ${\cal{D}}_{\bar{m}}$
 has a part which is even about the mirror planes $\bar{m}$, 
as may also be checked from Eq. (10). 
 
The above is fairly general. There is no reason why for a specific 
experimental geometry and ${\bf{p}}$,
       Eq. (10) may not be zero.
       The effect also has additional symmetries under rotation of ${\bf{p}}$.
       They depend on details of the current pattern in the proposed state
    and must be examined separately for each proposal.

 {\it Polarized ARPES for the Proposed State}:
We will now consider the special T-breaking states \cite{cmv2} predicted for under-doped cuprates.
Such states have been derived for a general Hamiltonian in the space of three orbitals 
per unit cell for non-local interactions above a critical value depending on the
deviation of electronic density $x$ away fron half-filling. The phase diagram 
in the $T-x$ plane, for the proposed T-breaking phase is consistent with the
observed "pseudo-gap" phase in the cuprates \cite{cmv3}.

 For the case that the
difference in energy of the
Cu-$d_{x^2 - y^2}$ level $\epsilon_d$ and the O-$p_{x,y}$ levels
$\epsilon_{\bf p}$ is much less than their hybridization energy $t_{pd}$
and for the direct Oxygen-Oxygen hopping parameter
$t_{pp} << t_{pd}$, the conduction band wave function
in a  tight-binding representation, $\it{without}$ T-breaking
may be written in terms
of the "anti-bonding" orbitals and the "non-bonding" orbitals as follows:
\begin{eqnarray}
| {\bf k} \rangle \simeq (N_k)^{-1}[a_{\bf k}^+ +
4\frac{t_{pp}}{\epsilon_{\bf{k}}} s_xs_y (s_x^2-s_y^2)n_{\bf k}^+ ]|0\rangle.
\end{eqnarray}
where the anti-bonding and the non-bonding orbitals are created respectively by
\begin{eqnarray}
a_{{\bf k}}^+ =
\frac{d_k^+}{\sqrt{2}} \: + \left(\:
\frac{s_x \, p_{kx}^+ + s_y \, p_{ky}^+}{\sqrt{2} s_{xy}}\right),~~
n_{{\bf k}}^+ = \left(s_y \: p_{kx}^+ - s_x \: p_{ky}^+ \right) / s_{xy}.
\end{eqnarray}
where $s_{x,y} = \sin (k_{x,y} a/2)$, $c_{x,y} = \cos (k_{x,y} a/2)$ and
$s_{xy}^2 = \sin^2\frac{ k_x a}{2} + \sin^2 \frac{k_y a}{2}$ and $\epsilon_{\bf{k}}
= 2t_{pd}s_{xy}$.
Spin labels have been suppressed. $d_k^+$, $p_{kx,y}^+$ are respectively the
 creation operators in momentum space for the $d_{x^2-y^2}$ atomic orbital at the
Cu-site $R_i$ and the $p_{x,y}$ orbitals at the oxygen site at
$( R_i + \frac{a_{x,y}}{2})$, in each cell $i$..

If one approximates $|{\bf{p}}\rangle$ by a plane wave $\sim exp(i{\bf p.r})$,
the difference of the current from rcp ARPES and lcp ARPES, Eq. (10)
 vanishes for all ${\bf{p}}$.
This was the result presented in Ref. (\cite{cmv3},\cite{error}). But for better 
outgoing wavefunctions
which include the lattice and surface potentials, Eq. (10) is in general finite,
except when ${\bf{p}}$ lies in the plane $\hat{m}$, as shown above.
For ${\bf{p}}$ about this condition,
 the difference is odd, again as shown above.

Two sets of wavefunction for a T-breaking phase preserving translational invariance
and inversion can be derived in the mean-field approximation \cite{cmv2}. The ground state
of $|\Theta_1\rangle$ is made up of products of $|{\bf{k}},\theta_1\rangle$ :
\begin{eqnarray}
|{\bf k},\theta_1 \rangle  & = & (N_k)^{-1}[a_{{\bf k},\theta_1}^+ +
4\frac{t_{pp}}{\epsilon_{\bf{k}}}
 s_xs_y (s_x^2-s_y^2)n_{\bf k}^+]|0\rangle   \label{thetaI} \\
\theta_1  & = & \pm\sum'_{\bf k} [s_x \langle p^{+}_{x{\bf k}}d_{\bf k}\rangle - s_y \langle p^{+}_{y,{\bf k}}d_{\bf k}\rangle]. \nonumber \\
a_{{\bf k},\theta_1}^+ & = &
\frac{d_k^+}{\sqrt{2}} \: + \left(\:
\frac{s_x(1+i\theta_1) \, p_{kx}^+ + s_y (1-i\theta_1)\, p_{ky}^+}{\sqrt{2} s_{xy}}\right),~~ \nonumber
\end{eqnarray}
The ground state of $|\Theta_2\rangle$ is made up of products of
\begin{eqnarray}
|{\bf k},\theta_2 \rangle  & = & (N_k)^{-1}[a_{{\bf k},\theta_2}^+ +
4\frac{t_{pp}} \epsilon_{\bf{k}} s_x s_y (s_x^2-s_y^2)
n_{\bf k}^+]|0\rangle \label{thetaII} \\
\theta_{2} & = & \pm\sum'_{\bf k}[c_x \langle p^{+}_{x{\bf k}}d_{\bf k}\rangle \pm c_y \langle p^{+}_{y{\bf k}}d_{{\bf k}}\rangle]. \nonumber \\
a_{{\bf k},\theta_2}^+ & = &
\frac{d_k^+}{\sqrt{2}} \: + \left(\:
\frac{(s_x+i\theta_2 c_x) \, p_{kx}^+ + (s_y \pm i\theta_2 c_y)\, p_{ky}^+}{\sqrt{2} s_{xy}}\right),~~ \nonumber
\end{eqnarray}

In (\ref{thetaI},\ref{thetaII}), the expectation values are determined self-consistently and
$(\theta_1, \theta_{2}) << 1$ are assumed. 
The derived additional terms, proportional to the $\theta$'s break T-invariance
because the effective Hamiltonians, of which Eqs. (\ref{thetaI},\ref{thetaII}) are eigenstates, 
cannot be made real by any unitary transformation. The ground state currents corresponding to
$|\Theta_1\rangle$ and $|\Theta_2\rangle$ are shown in Figs. (1a) and (1b) respectively.
 $|\Theta_1\rangle$ 
 retains mirror symmetry about the $x=0,y=0$
planes, but not about the mirror planes $(\bar{m}_1 :x=\pm y)$.
On the other hand $|\Theta_{II}\rangle$ 
 does not retain mirror symmetry about  the mirror planes $(\bar{m}_2$ :$x=0$ and $y=0$).
 Two of the 4 possible domains of $|\Theta_2\rangle$ retain reflection symmetry about 
$x=y$ but not about $x=-y$ while the other two have the opposite behavior.
However $\cal{D}$ can be shown to be zero in $|\Theta_2\rangle$ for all of them at both
$k_x = \pm k_y$ due to the symmetry of the transfer integral among the two
 oxygen orbitals in each unit cell. 

The symmetry of the states (\ref{thetaI},\ref{thetaII}) has the 
following consequence for $\cal{D}$.
   The state $|\Theta_1\rangle$  produces
an effect in ${\cal{D}}$ of order $\theta_1$ which is even 
about the $x=\pm y$ mirror planes and zero effect at the $x=0,y=0$ mirror planes. 
The state $|\Theta_2\rangle$ 
produce an effect in $\cal{D}$ of order $\theta_{2}$ which is even about the
mirror planes $x=0$ and $y=0$. From Eq. (10) it follows that the effect changes sign
at these two mirror planes (i.e. if it is positive at one, it is negative at the other)
and have maximum absolute magnitude at $(k_xa,k_ya) = (\pm \pi,0)$. 
The effect is zero at the mirror planes
$x = \pm y$. 

Togather with the geometric effect, the {\it effective} mirror planes defined as the
plane for
${\cal{D}} = 0$ 
 therefore appear rotated compared to the
{\it crystalline} mirror planes; The rotation is in opposite directions for two mutually
orthogonal crystalline mirror planes $\bar{m}$. Further $\bar{m}$ are the $x = \pm y$ planes 
for the state $|\Theta_{I}\rangle$ and the planes $x = 0$ and $y = 0$ for the state
$|\Theta_{II}\rangle$.   

There are additional modifications of the wave-functions near the chemical potential
in the CC phase derived in \cite{cmv2}, which are not included above.

{\it Analysis of the Experiments}:
Recent polarized ARPES experiments \cite{kaminski} to
look for the predicted effect \cite{cmv3} give results which are
consistent  with  T-breaking 
in the underdoped phase of the cuprates. In one set of experiments \cite{kaminski}, 
the region
of momentum at the edge of the first Brillouin zone near the point $(\pi /a,0)$  was
investigated thoroughly with $\hat{n}$ normal to the $Cu-O$ plane. 
In the absence of a pseudogap the difference in the
current for rcp and lcp ARPES was found to be odd about this point
 in traveling along the
edge of the zone. This serves as a check on the experimental set up.
In underdoped samples with pseudogap, a difference,
symmetric about this
 point was observed below the temperature of appearance of the pseudogap and none was 
seen above this temperatures. This result has been seen in several underdoped
samples; overdoped samples do not show the effect.

Two other features of the results are especially noteworthy in relation to microscopic theory.
 An investigation  of the Brillouin zone near
$(0,\pi/a)$ in a given crystal produced an effect in $\cal{D}$ of opposite sign
to that around $(\pi/a,0)$ \cite{kaminski}. In other words the mirror planes $x=0$
and $y=0$ are ${\it effectively}$ rotated in opposite directions. Therefore according 
to the symmmetry considerations above these
 experiments are consistent 
with the state
 $|\Theta_{II}\rangle$. Secondly, the magnitude
 of the effect is independent
of the energy in the range investigated, $\sim 0.5 eV$. This is important because 
in the microscopic theory \cite{cmv2},
T-breaking is produced in a three-band model, by admixing due to long-range interactions,
the states of the
 conduction and valence bands of the non-interacting Hamiltonian. The magnitude of the effect 
is then essentially uniform over the entire conduction band. (This is not a Fermi-surface effect).
 Moreover the absolute
 magnitude of the effect, about $5\%$, is consistent with the expectations.

In another set of experiments \cite{golden}, the condition that
  the $\hat{n}$ is in the mirror plane  was followed only for investigation with
outgoing electron momentum $\bf{p}$ in the $x=y$ mirror plane.
 Then, within the experimental errors
no difference was observed between rcp and lcp ARPES when 
the point at the Fermi-surface in the ($k_x=k_y$) direction was investigated.
This is also consistent with the proposed CC state $|\Theta_{II}\rangle$.

We now inquire if any other symmetry breaking can produce the observed effects.
It may be seen that lattice distortions of the tetragonal to orthorhombic type,
while preserving translational symmetry, do not have the symmetry to produce 
the observed effect. It may in principle be produced by  distortions of the basic two-dimensional
unit cell to the shape of a 
parallelogram, so that the relevant mirror plane symmetries are lost. 
Careful investigation of lattice distortions as a function of temperature \cite
{kaminski} have not revealed
any nor have such distortions have been reported elsewhere.

The difference of ARPES intensity for rcp and lcp photons exists  for
 any T-breaking phase, be it due to spin-order or orbital order. Antiferromagnetic order
at ${\bf Q}=(\pi/a,\pi/a)$ would produce a phase breaking reflection symmetry about the
$x=\pm y$ planes and is therefore incompatible with the observations. It would produce 
zero-effect for outgoing momenta along the $\bar{m}$ planes $x=0, y=0$ and maximum at
the planes $x=\pm y$. Further the effect would be zero at the zone-edge in these directions.
A proposal combining the staggered flux phase \cite{staggered}, 
with the idea
 of a Quantum critical point near optimum doping has recently been advanced as a
possibility for the underdoped compounds with the name, "D-density wave" \cite{chakra}.
This phase has the same symmetry as the above antiferromagnetic phase with regard to the
ARPES experiment.
Moreover both these phases break translational symmetry so that the Fermi-surface 
 consists of four pockets in the $\pi -\pi$ directions. This is contrary to 
the observations by ordinary (linearly polarized ) ARPES in the cuprates. 
More complicated magnetic or structural symmetry-breakings can be envisaged
 producing the observed effect. But they would have to have escaped notice in direct
diffraction experiments.

We also note that an ${\it Anyon}$ state \cite{laughlin} can also be detected by 
ARPES experiments. For such a state the effective rotation of all the crystalline 
planes would be in the same direction.

{\it Conclusions}: The existence of a quantum critical point in the phase diagram\cite{schon}
of Copper-oxides near optimum doping suggested that the central feature to be understood 
in the cuprates is the symmetry of the underdoped state. In the ARPES experiments,
 an effect has been discovered which purely on symmetry grounds (as well as on grounds
of its energy independence) has been identified here to
be consistent with a proposed state \cite{cmv2}. Such a state breaking Time-reversal invariance
but not translational invariance is the property of a general copper-oxide model with 
long range interactions.
It is of-course obvious that a state which does not break translational invariance but has 
ground state currents can only be the property of a model with atleast three orbitals 
per unit cell.
Approximate solutions of the same model lead to essentially all \cite{unexplained}
 the universal features
of the phase diagram of the cuprates, including the marginal Fermi-liquid fluctuations 
in the normal state 
near optimum doping, the pseudogap phenomena at underdoping, the 
crossover to a Fermi-liquid at overdoping, the vertex for "d-wave" pairing 
as well as the right energy scale and
coupling constant for the high $T_c$ of the cuprates.

{\it Acknowledgements:}We have benefitted enormously from discussions with A.Kaminsky and 
J.C.Campuzano on the
 experimental results at various stages of their experiment. 
and from discussions with
Elihu Abrahams, Eugene Blount and Ashvin Vishwanath about 
the theoretical issues.


\begin{references}
\bibitem{footnote}
In a crystal T-breaking preserving all the reflection symmetries
 and point group symmetries
 can occur only if the elementary magnetic moment (due to orbital currents or spins)
is confined to the irreducible volume of the unit cell.
\bibitem{error}
There are several other errors of algebra in the calculations of Ref. \cite{cmv3}
which led to the results for ARPES experiment for the $|\Theta_I\rangle$ phase which actually
pertain to the $|\Theta_{II}\rangle$ phase.
\bibitem{unexplained}
One feature which is not explained is why the occurence of the T-breaking phase is 
unaccompanied by thermodynamic signatures of a phase transition.
\end{references}
      \end{document}